\documentclass[12pt, letterpage]{article}
\usepackage{amssymb,latexsym}
\usepackage{natbib}
\usepackage[unicode=true, bookmarks=true, bookmarksnumbered=false, bookmarksopen=false, breaklinks=true, pdfborder={0 0 0},backref=false]{hyperref}
\hypersetup{
  colorlinks=true,
  allcolors=blue
}

\usepackage{appendix}
\usepackage{amsmath}
\usepackage{bbold}
\usepackage{graphicx}
\usepackage{cases}
\usepackage{float}
\usepackage{subcaption}
\usepackage{multirow}
\usepackage{xcolor}
\usepackage{indentfirst,geometry,setspace,fancyhdr,sectsty}

\allowdisplaybreaks

\def\Pr{\hbox{Pr}}

\begin{document}
\title{When Good News Fails in Evolution:\\
{\large\textit{A Note on Information, Exposure, and Evolutionary Stability}}}
\author{Jiabin Wu\thanks{Department of Economics, University of Oregon, jwu5@uoregon.edu.}}
\maketitle
\begin{abstract}
Good news about a strategy need not translate into evolutionary success. We study how the interaction between information and strategy exposure shapes evolutionary dynamics and equilibrium predictions. We derive the associated population dynamic from a simple revision process in which agents encounter candidate strategies at state-dependent rates and adopt them after receiving favorable payoff signals. We distinguish the monotone likelihood ratio property (MLRP) of signals from effective MLRP, which accounts for exposure relative to population shares and determines whether better-performing strategies grow faster. Random population encounters preserve MLRP and generate payoff-monotone evolution. When exposure sufficiently favors rare strategies, effective MLRP is violated near strict pure Nash equilibria, rendering them unstable. Conversely, when exposure sufficiently favors popular strategies, effective MLRP is violated near non-Nash pure states, allowing them to remain locally stable. Thus, even when signals satisfy MLRP, differences in strategy exposure can reverse their evolutionary implications and alter equilibrium predictions.
\end{abstract}
\newpage
\section{Introduction}
Evolutionary game theory has developed a wide range of dynamics to explain how payoff differences shape the evolution of behavior \citep{Weibull1995, Sandholm2010}. A common principle underlying many of these dynamics is that better-performing strategies are more likely to be adopted when considered. Yet the resulting evolutionary predictions can differ substantially, depending on how agents encounter and evaluate competing strategies. This raises a broader question that has long been studied in evolutionary game theory: under what conditions does an advantage in payoff information translate into faster population growth, and how does this affect equilibrium stability?

We revisit this question from a different angle by treating strategy exposure and the information generated upon evaluation as separate primitives. To formalize this distinction, we consider a learning process in which agents encounter strategies, receive signals about their payoffs, and adopt a strategy upon receiving a favorable signal. Following \cite{Milgrom1981}, we characterize the informativeness of these signals using the monotone likelihood ratio property (MLRP). We introduce general exposure functions that determine how frequently strategies are encountered as a function of the population state. This framework accommodates uniform encounters across strategies, random population encounters, and exposure biased toward relatively rare or popular strategies. It also allows the exposure of a strategy to depend on the population shares of competing strategies. A familiar illustration is the contrast between the logit dynamic \citep{Blume1993,FudenbergLevine1998} and the imitative logit dynamic \citep{Weibull1995,BjornerstedtWeibull1996}. Both can be represented by the same signal process, in which the probability of receiving a favorable signal increases exponentially with a strategy's payoff. The difference lies in exposure: under logit, strategies are encountered uniformly, whereas under imitative logit, they are encountered in proportion to their population shares.

Our analysis distinguishes between MLRP of the original signals and what we call effective MLRP. While the former captures whether favorable signals are more likely to come from better-performing strategies, the latter captures whether this informational advantage survives the sampling process and translates into faster relative population growth. We show that random population encounters preserve MLRP and generate payoff-monotone evolution. Under sufficiently strong exposure biases favoring rare strategies, effective MLRP is violated near strict pure Nash equilibria, which can become unstable. Conversely, exposure biases favoring popular strategies can violate effective MLRP near non-Nash pure states, allowing them to remain locally stable. Thus, even when signals consistently provide good news about better-performing strategies, the way strategies are encountered can prevent this information from translating into evolutionary success.

\subsection*{Literature Review}

The broader literature on information and social learning also emphasizes how the structure of communication and sampling shapes collective outcomes. In \cite{EllisonFudenberg1995}, agents learn from the choices and realized payoffs of randomly sampled individuals. Because an alternative must appear in the sample to be adopted, the sampling process creates an implicit popularity bias that affects whether superior alternatives spread. \cite{BanerjeeFudenberg2004} study Bayesian learning in which agents infer the relative quality of alternatives from both their observed frequencies in a sample and noisy signals about their payoffs. They also examine how biased sampling affects these inferences and long-run outcomes. In cultural evolution, \cite{VidiellaEtAl2022} analyze how information transparency interacts with popularity bias, showing that strong conformity can impede the spread of beneficial innovations. These papers demonstrate that the effects of information depend on the social processes through which individuals encounter and learn about alternatives. Our focus is on the implications of this interaction for payoff monotonicity and evolutionary stability.

More specifically, our analysis relates to the informational foundations of evolutionary dynamics. \cite{Sandholm2010} classifies revision protocols according to their information requirements and distinguishes imitative protocols, in which candidate strategies are identified through population encounters, from direct protocols, in which alternatives can be considered independently of their current use. He establishes how these protocols differ in their ability to generate aggregate behavior consistent with payoff incentives and Nash equilibrium. Our analysis complements this approach by examining how the informativeness of signals and the exposure of candidate strategies jointly shape evolutionary outcomes.

The sampling best-response literature studies how agents respond to observed actions \citep{Sandholm2001,OyamaSandholmTercieux2015} or draw statistical inferences from them \citep{SalantCherry2020,SawaWu2023,IzquierdoIzquierdo2025}. \cite{ArieliArigapudi2024} combine sampled choices with noisy signals about product quality. Related work on experience-based evolutionary dynamics examines how agents evaluate strategies using payoffs from sampled trials or actual interactions \citep{OsborneRubinstein1998,Sethi2000,SandholmIzquierdoIzquierdo2020,IzquierdoIzquierdo2022,ArigapudiHellerMilchtaich2021,ArigapudiHeller2025,Wu2027}. Rather than analyzing particular sampling or payoff-evaluation procedures, we identify general properties of strategy exposure that determine whether better-performing strategies enjoy an evolutionary advantage.

\section{The Model}

Consider a continuum of agents with a common finite strategy set \(S=\{1,2,\ldots,N\}\). Let \(x=(x_1,\ldots,x_N)^T\) denote the population state, where \(x_i\) is the proportion of agents playing strategy \(i\). The state space is \(X=\{x\in\mathbb{R}_+^N:\sum_{i=1}^N x_i=1\}\). Let \(F:X\to\mathbb{R}^N\) be a continuous payoff function, where \(F_i(x)\) is the payoff to strategy \(i\) at population state \(x\). Let \(e_i\) denote the monomorphic state in which every agent plays strategy \(i\).

Each agent is equipped with an independent Poisson alarm clock with rate one. When the clock rings, the agent receives an opportunity to revise her strategy. Revision involves two distinct steps: encountering a candidate strategy and evaluating it. An evaluation generates either good news ($G$) or bad news ($B$) about the candidate's payoff. The agent adopts a candidate upon receiving good news. We first describe the information generated by an evaluation and then the process through which candidate strategies are encountered.

Suppose that evaluating strategy $i$ at state $x$ produces good news with probability
\begin{equation}
 a_i(x)=\Pr(G\mid i,x)=a(F_i(x)),
 \qquad \Pr(B\mid i,x)=1-a(F_i(x)),
\end{equation}
where $a$ is a continuous function taking values strictly between zero and one. The same function applies to all strategies. Following \cite{Milgrom1981}, we assume that the signals satisfy the monotone likelihood ratio property (MLRP): for any $u'>u$,
\begin{equation}
 \frac{a(u')}{1-a(u')}\geq \frac{a(u)}{1-a(u)}.
\end{equation}
For binary signals, MLRP is equivalent to $a$ being nondecreasing. Thus, a better-performing strategy is at least as likely to generate good news when evaluated. We use strict MLRP when a strict ranking is needed. Agents need not observe expected payoffs or the population state; they respond only to the signals received upon evaluation.

We next specify how agents encounter candidate strategies. At each population state $x$, opportunities to encounter strategy $i$ arrive according to an independent Poisson process with rate $g_i(x)\geq 0$. These encounter rates capture how frequently different strategies become available for evaluation and may depend on the entire population state. We assume that $g_i$ is continuous for every $i$ and that $\sum_{j=1}^N g_j(x)>0$ for all $x\in X$.

When an agent searches for a candidate strategy, she evaluates the first strategy encountered. Since the waiting times for encounters are independently exponentially distributed, the probability that strategy $i$ is encountered first is
\begin{equation}
m_i(x)=\frac{g_i(x)}{\sum_{j=1}^N g_j(x)}.
\label{eq:exposure}
\end{equation}
We refer to $m(x)=(m_1(x),\ldots,m_N(x))$ as the \emph{strategic exposure function}. It determines the probability that each strategy is evaluated, independently of the information generated by that evaluation. The encounter rates are properties of the search environment; agents need not know the population distribution or calculate these rates.

This formulation accommodates several familiar exposure procedures. Under uniform exposure, $g_i(x)=1$ for every $i$, so $m_i(x)=1/N$. Under random population encounters, each individual is equally likely to be encountered, implying $g_i(x)=x_i$ and hence $m_i(x)=x_i$.

More generally, consider encounter rates of the form
\begin{equation}
g_i(x)=x_i^\alpha,\qquad \alpha>0.
\end{equation}
The resulting exposure function is
\begin{equation}
m_i(x)=\frac{x_i^\alpha}{\sum_{j=1}^N x_j^\alpha}.
\label{eq:powerexposure}
\end{equation}
When $\alpha=1$, this reduces to random population encounters. When $0<\alpha<1$, relatively rare strategies receive greater exposure relative to their population shares, whereas $\alpha>1$ favors relatively popular strategies. Uniform exposure corresponds to the limiting case $\alpha=0$ at interior states, or equivalently to setting $g_i(x)=1$ for all states. 

Note that the general encounter rates may depend on the shares of all competing strategies. In what follows, we take the strategic exposure function $m:X\to X$ as a primitive and allow for general functional forms.

We now derive the revision protocol. At a revision opportunity, an agent draws candidate strategies according to the exposure function $m(x)$ and evaluates them in succession until one generates good news. Conditional on the population state, draws and signals are independent across attempts. Let
\begin{equation}
 D(x)=\sum_{j=1}^N m_j(x)a_j(x)>0
\end{equation}
denote the probability that an evaluation generates good news. The probability that the agent eventually adopts strategy $i$ is
\begin{equation}
 \begin{aligned}
 \rho_i(x)
 &=\sum_{r=0}^{\infty}(1-D(x))^r m_i(x)a_i(x)=\frac{m_i(x)a_i(x)}{\sum_{j=1}^N m_j(x)a_j(x)}.
 \end{aligned}
 \label{eq:revision}
\end{equation}
Since this probability does not depend on the agent's current strategy, the probability of switching from $j$ to $i\neq j$ at a revision opportunity is $\rho_i(x)$. The associated mean dynamic is therefore
\begin{equation}
 \begin{aligned}
 \dot{x}_i
 &=\sum_{j\neq i}x_j\rho_i(x)-x_i\sum_{j\neq i}\rho_j(x)=\frac{m_i(x)a_i(x)}{\sum_{j=1}^N m_j(x)a_j(x)}-x_i,
 \qquad i\in S.
 \end{aligned}
 \label{eq:dynamic}
\end{equation}
We refer to \eqref{eq:dynamic} as the \emph{exposure-based evolutionary dynamic}.\footnote{If an agent instead makes only one evaluation at each revision opportunity and retains her strategy following bad news, the mean dynamic becomes $\dot{x}_i=m_i(x)a_i(x)-x_iD(x)$. It differs from \eqref{eq:dynamic} only by the strictly positive factor $D(x)$ and thus has the same trajectories up to a change of speed.}

To relate this dynamic to familiar examples, suppose that $a(u)=c\exp(u/\eta)$, where $\eta>0$ and $c>0$ is sufficiently small that $a(u)<1$ for every feasible payoff. Under uniform exposure, \eqref{eq:dynamic} becomes the logit dynamic \citep{Blume1993,FudenbergLevine1998}:
\begin{equation}
 \dot{x}_i=\frac{\exp(F_i(x)/\eta)}{\sum_{j=1}^N\exp(F_j(x)/\eta)}-x_i.
\end{equation}
Under random population encounters, the same signal structure generates the imitative logit dynamic \citep{Weibull1995,BjornerstedtWeibull1996}:
\begin{equation}
 \dot{x}_i=\frac{x_i\exp(F_i(x)/\eta)}{\sum_{j=1}^N x_j\exp(F_j(x)/\eta)}-x_i.
\end{equation}
The distinction between the two dynamics lies entirely in how the candidate strategies are encountered.

%Finally, for strategies $i$ and $j$ that are both in use, \eqref{eq:dynamic} implies
%\begin{equation}
% \frac{\dot{x}_i}{x_i}-\frac{\dot{x}_j}{x_j}
 %=\frac{1}{D(x)}\left(\frac{m_i(x)}{x_i}a_i(x)
 %-\frac{m_j(x)}{x_j}a_j(x)\right).
 %\label{eq:relativegrowth}
%\end{equation}
%Thus, the ranking of percentage growth rates depends not only on the probabilities of good news but also on strategy exposure relative to population shares. This is the distinction underlying the analysis that follows.

\section{Analysis}

We now examine how the information generated by evaluating strategies interacts with their exposure to determine evolutionary outcomes. The central distinction is between the likelihood of receiving good news about a strategy and the frequency with which the strategy is encountered. While MLRP restricts the former, it places no restriction on the latter.

\subsection{Effective MLRP and payoff monotonicity}

Consider two strategies $i$ and $j$ with positive population shares. From \eqref{eq:dynamic}, their percentage growth rates satisfy
\begin{equation}
 \frac{\dot{x}_i}{x_i}=\frac{m_i(x)}{x_i}\frac{a_i(x)}{D(x)}-1,
 \qquad
 \frac{\dot{x}_j}{x_j}=\frac{m_j(x)}{x_j}\frac{a_j(x)}{D(x)}-1,
 \label{eq:percentagegrowth}
\end{equation}
where $D(x)=\sum_{k=1}^N m_k(x)a_k(x)$. Consequently,
\begin{equation}
 \frac{d}{dt}\log\frac{x_i}{x_j}
 =\frac{1}{D(x)}\left[
 \frac{m_i(x)}{x_i}a_i(x)-\frac{m_j(x)}{x_j}a_j(x)
 \right].
 \label{eq:loggrowth}
\end{equation}
The change in the relative population shares thus depends on the probability of good news multiplied by exposure per current user of each strategy. MLRP by itself compares only the probabilities of good news.

To separate these two forces, suppose that strategies with positive population shares also have positive exposure. The logarithm of the ratio of the two exposure-adjusted good-news probabilities can be written as
\begin{equation}
 \begin{aligned}
 \log\frac{m_i(x)a_i(x)/x_i}{m_j(x)a_j(x)/x_j}
 &=\underbrace{\log\frac{a_i(x)}{a_j(x)}}_{\text{payoff information}}
 +\underbrace{\log\frac{m_i(x)/x_i}{m_j(x)/x_j}}_{\text{relative exposure}}=\log\frac{a_i(x)}{a_j(x)}
 +\log\frac{g_i(x)/x_i}{g_j(x)/x_j}.
 \end{aligned}
 \label{eq:logdecomposition}
\end{equation}
The second equality follows from $m_i(x)=g_i(x)/\sum_k g_k(x)$. By \eqref{eq:loggrowth}, the sign of \eqref{eq:logdecomposition} determines which strategy has the higher percentage growth rate.

If $F_i(x)>F_j(x)$, MLRP implies $a_i(x)\geq a_j(x)$, so the first term in \eqref{eq:logdecomposition} is nonnegative. The second term, however, can have either sign. Exposure favoring strategy $j$ sufficiently strongly can therefore give it a higher percentage growth rate even when its expected payoff is lower.

\paragraph{Definition 1 (Effective MLRP).}
At a population state $x$, the exposure and signal processes satisfy \emph{effective MLRP} if, for every pair of strategies $i,j$ with $x_i,x_j>0$,
\begin{equation}
 \begin{aligned}
 F_i(x)>F_j(x)&\ \Longrightarrow\
 \frac{m_i(x)}{x_i}a_i(x)\geq \frac{m_j(x)}{x_j}a_j(x),\\
 F_i(x)=F_j(x)&\ \Longrightarrow\
 \frac{m_i(x)}{x_i}a_i(x)=\frac{m_j(x)}{x_j}a_j(x).
 \end{aligned}
 \label{eq:effectivemlrp}
\end{equation}
Effective MLRP is \emph{strict} if the first inequality is strict whenever the two payoffs differ. We say that effective MLRP holds for the dynamic if it holds at every population state for all strategies in its support.

The definition concerns the combined effects of payoff information and exposure. In contrast to MLRP of the original signals, effective MLRP requires the information advantage of a better-performing strategy to survive the encounter process. By \eqref{eq:loggrowth}, effective MLRP implies weak payoff monotonicity, while strict effective MLRP implies strict payoff monotonicity.

Under random population encounters, $m_i(x)=x_i$, and hence $m_i(x)/x_i=1$ for each strategy in use. Effective MLRP then follows directly from MLRP of the original signals. With general exposure, the relative-exposure term in \eqref{eq:logdecomposition} need not vanish, and effective MLRP may fail even though the original signals satisfy strict MLRP.

\subsection{Stability of monomorphic states}

We next study the implications for the stability of pure population states. Recall that $e_i$ denotes the state in which every agent plays strategy $i$. We call $e_i$ a Nash equilibrium if $F_i(e_i)\geq F_j(e_i)$ for every $j\in S$, and a strict Nash equilibrium if $F_i(e_i)> F_j(e_i)$ for every $j \ne i$. A non-Nash pure state is one at which some alternative strategy earns a strictly higher payoff.

Since $a_j(x)>0$ for every strategy, $e_i$ is a rest point of \eqref{eq:dynamic} if and only if $m_i(e_i)=1$. Thus, a pure state cannot be stationary if strategies absent from the population continue to receive positive exposure. In particular, none of the pure states is stationary under uniform exposure. For the remainder of this subsection, we consider pure states satisfying $m_i(e_i)=1$.

For a strategy $j\neq i$ with $x_j>0$, its percentage growth rate near $e_i$ is
\begin{equation}
 \frac{\dot{x}_j}{x_j}
 =\frac{m_j(x)}{x_j}\frac{a_j(x)}{D(x)}-1.
 \label{eq:mutantgrowth}
\end{equation}
By continuity, $D(x)\to a_i(e_i)$ and $a_j(x)\to a_j(e_i)$ as $x\to e_i$. Thus, the critical level of exposure per current user of the rare strategy $j$ is
\begin{equation}
 \frac{a_i(e_i)}{a_j(e_i)}.
 \label{eq:exposurethreshold}
\end{equation}
When $m_j(x)/x_j$ remains below this threshold sufficiently close to $e_i$, strategy $j$ declines; when it remains above the threshold, strategy $j$ grows. This comparison yields the following results.

\paragraph{Proposition 1 (Strict Nash equilibria).} 
Suppose $e_i$ is a strict Nash equilibrium, $m_i(e_i)=1$, and the signals satisfy strict MLRP at the payoffs $F_i(e_i)$ and $F_j(e_i)$, $j\neq i$.
\begin{enumerate}
 \item If, for every $j\neq i$,
 \begin{equation}
 \limsup_{\substack{x\to e_i\\x_j>0}}
 \frac{m_j(x)}{x_j}
 <\frac{a_i(e_i)}{a_j(e_i)},
 \label{eq:nashstable}
 \end{equation}
 then $e_i$ is asymptotically stable.
 \item If, for some $j\neq i$,
 \begin{equation}
 \liminf_{\substack{x\to e_i\\x_j>0}}
 \frac{m_j(x)}{x_j}
 >\frac{a_i(e_i)}{a_j(e_i)},
 \label{eq:nashunstable}
 \end{equation}
 then $e_i$ is unstable.
\end{enumerate}

\noindent\emph{Proof.}
Under strict MLRP and strict Nash equilibrium, $a_i(e_i)>a_j(e_i)$ for every $j\neq i$. If \eqref{eq:nashstable} holds, continuity of $a_j$ and $D$ implies that there exist a neighborhood $U$ of $e_i$ and $\varepsilon>0$ such that, for every $j\neq i$ and $x\in U$ with $x_j>0$, equation \eqref{eq:mutantgrowth} yields $\dot{x}_j\leq-\varepsilon x_j$. The bound in \eqref{eq:nashstable} and continuity of $m_j$ also imply $m_j(x)=0$ whenever $x_j=0$ in a sufficiently small neighborhood, so the same inequality holds there. Hence, writing $1-x_i=\sum_{j\neq i}x_j$, we obtain
\[
 \frac{d}{dt}(1-x_i)=\sum_{j\neq i}\dot{x}_j
 \leq-\varepsilon(1-x_i)
\]
throughout that neighborhood. It follows that $e_i$ is asymptotically stable.

If \eqref{eq:nashunstable} holds, continuity instead gives a neighborhood of $e_i$ and $\varepsilon>0$ such that $\dot{x}_j\geq\varepsilon x_j$ whenever $x_j>0$ in that neighborhood. Starting arbitrarily close to $e_i$ with a positive share of strategy $j$, this share grows until the trajectory leaves the neighborhood. Therefore $e_i$ is unstable. \hfill$\square$
\\

Proposition 1 shows why an informational advantage need not protect a strict Nash equilibrium. At $e_i$, strict MLRP favors the incumbent strategy because $a_i(e_i)>a_j(e_i)$ for every alternative $j$. But a sufficiently strong exposure advantage for rare strategies can more than offset this difference. Under random population encounters, $m_j(x)/x_j=1$, so the first condition holds and every strict Nash equilibrium is asymptotically stable. In contrast, if exposure per current user of any rare strategy grows without bound near $e_i$, the second condition holds and the strict Nash equilibrium is unstable.

The reverse phenomenon can arise at a pure state that is not a Nash equilibrium.

\paragraph{Proposition 2 (Non-Nash pure states).}
Suppose $e_i$ is a non-Nash pure state and $m_i(e_i)=1$.
\begin{enumerate}
 \item If there is a strategy $j$ with $F_j(e_i)>F_i(e_i)$ such that
 \begin{equation}
 \liminf_{\substack{x\to e_i\\x_j>0}}
 \frac{m_j(x)}{x_j}
 >\frac{a_i(e_i)}{a_j(e_i)},
 \label{eq:nonnashunstable}
 \end{equation}
 then $e_i$ is unstable.
 \item If, for every $j\neq i$,
 \begin{equation}
 \limsup_{\substack{x\to e_i\\x_j>0}}
 \frac{m_j(x)}{x_j}
 <\frac{a_i(e_i)}{a_j(e_i)},
 \label{eq:nonnashstable}
 \end{equation}
 then $e_i$ is asymptotically stable.
\end{enumerate}

\noindent\emph{Proof.}
The argument follows from \eqref{eq:mutantgrowth}, as in the proof of Proposition 1. Condition \eqref{eq:nonnashunstable} makes the percentage growth rate of strategy $j$ strictly positive in a neighborhood of $e_i$, proving instability. Under \eqref{eq:nonnashstable}, every alternative strategy has a strictly negative percentage growth rate near $e_i$, so their total population share converges to zero and $e_i$ is asymptotically stable. \hfill$\square$
\\

Under strict MLRP, a profitable alternative $j$ satisfies $a_j(e_i)>a_i(e_i)$, so the threshold in \eqref{eq:nonnashunstable} is less than one. Thus, random population encounters make every non-Nash pure state unstable. Nevertheless, a sufficiently strong exposure bias toward the incumbent strategy can keep all alternative strategies below their respective thresholds, making a non-Nash pure state asymptotically stable. In this case, better-performing alternatives generate more good news when evaluated, but they are encountered too infrequently to expand.

 We illustrate these propositions using the power-exposure specification in \eqref{eq:powerexposure}, which is also employed by \cite{VidiellaEtAl2022} in their model of cultural evolution. For any $j\neq i$ with $x_j>0$,
\begin{equation}
 \frac{m_j(x)}{x_j}
 =\frac{x_j^{\alpha-1}}{\sum_{k=1}^N x_k^\alpha}
 \longrightarrow
 \begin{cases}
 +\infty,&0<\alpha<1,\\
 1,&\alpha=1,\\
 0,&\alpha>1,
 \end{cases}
 \qquad\text{as }x\to e_i.
 \label{eq:powerlimit}
\end{equation}

\paragraph{Corollary 1 (Power exposure).}
Suppose exposure is given by \eqref{eq:powerexposure} and the original signals satisfy strict MLRP.
\begin{enumerate}
 \item If $0<\alpha<1$, every pure state is unstable, including strict Nash equilibria.
 \item If $\alpha=1$, every strict Nash equilibrium is asymptotically stable and every non-Nash pure state is unstable.
 \item If $\alpha>1$, every pure state is asymptotically stable, including non-Nash pure states.
\end{enumerate}

\noindent\emph{Proof.}
The three conclusions follow immediately from \eqref{eq:powerlimit} and Propositions 1 and 2. In the first case, positive exposure per current user of every rare alternative diverges. In the second case, relative exposure equals one and strict MLRP determines the direction of growth. In the third case, exposure per current user of every rare alternative converges to zero. \hfill$\square$
\\

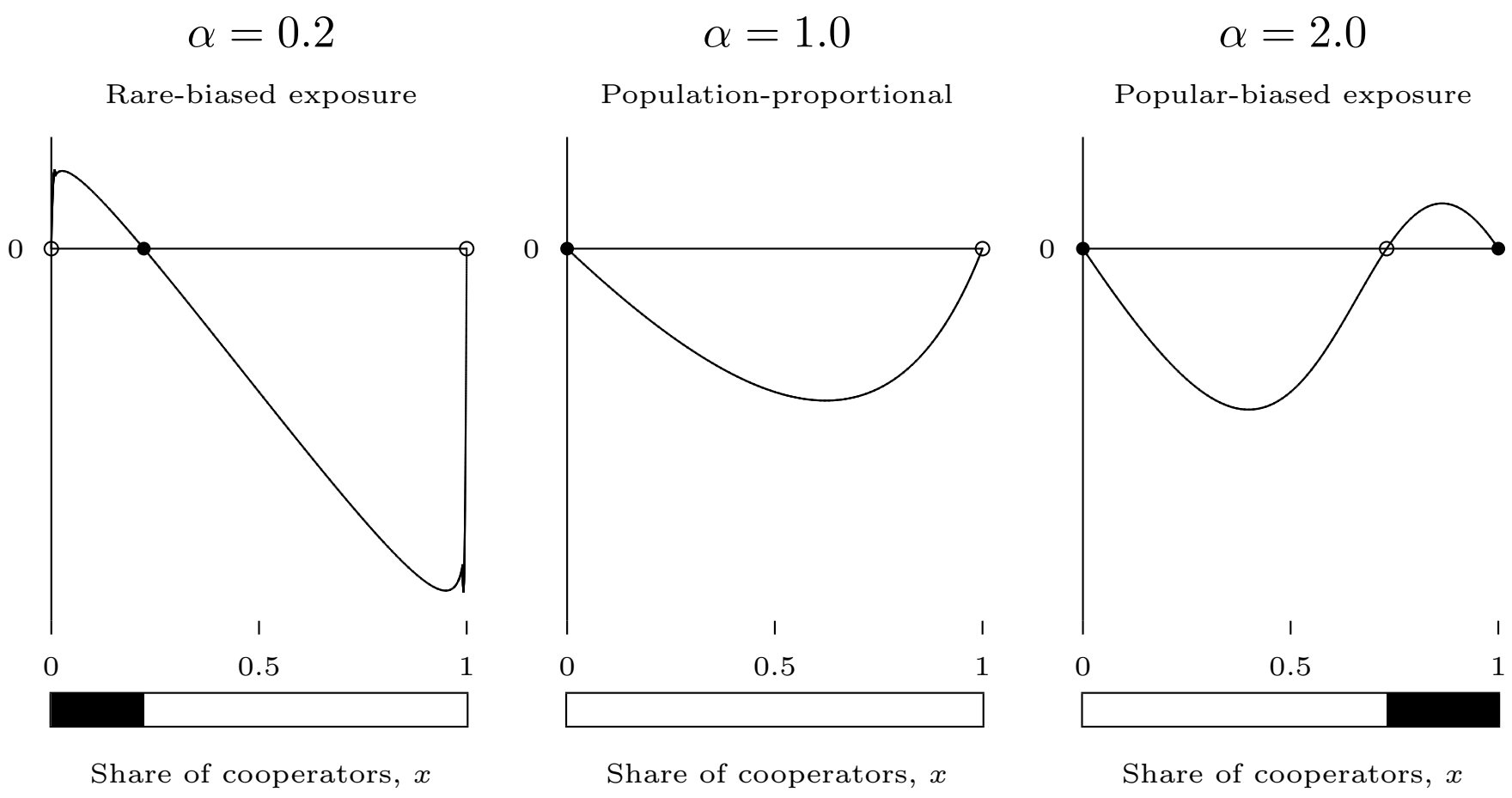
\begin{figure}[!ht]
\centering
\setlength{\unitlength}{1pt}
\begin{picture}(336,185)
  % Panel: alpha=0.2
  \put(60.000,175){\makebox(0,0){\footnotesize $\alpha=0.2$}}
  \put(60.000,161){\makebox(0,0){\tiny Rare-biased exposure}}
  \put(14.000,46.000){\line(0,1){106}}
  \put(14.000,127.538){\line(1,0){91}}
  \put(14.000,46){\line(0,-1){3}}
  \put(14.000,36){\makebox(0,0){\tiny 0}}
  \put(59.500,46){\line(0,-1){3}}
  \put(59.500,36){\makebox(0,0){\tiny 0.5}}
  \put(105.000,46){\line(0,-1){3}}
  \put(105.000,36){\makebox(0,0){\tiny 1}}
  \put(8.000,127.538){\makebox(0,0)[r]{\tiny $0$}}
  \put(14.000,23){\framebox(91,7){}}
  \put(14.000,23){\rule{20.266pt}{7pt}}
  \qbezier(14.000,127.538)(14.455,148.925)(14.910,143.570)
  \qbezier(14.910,143.570)(15.365,144.335)(15.820,144.458)
  \qbezier(15.820,144.458)(16.275,144.619)(16.730,144.538)
  \qbezier(16.730,144.538)(17.185,144.464)(17.640,144.262)
  \qbezier(17.640,144.262)(18.095,144.063)(18.550,143.784)
  \qbezier(18.550,143.784)(19.005,143.505)(19.460,143.172)
  \qbezier(19.460,143.172)(19.915,142.840)(20.370,142.467)
  \qbezier(20.370,142.467)(20.825,142.095)(21.280,141.692)
  \qbezier(21.280,141.692)(21.735,141.290)(22.190,140.863)
  \qbezier(22.190,140.863)(22.645,140.437)(23.100,139.991)
  \qbezier(23.100,139.991)(23.555,139.545)(24.010,139.083)
  \qbezier(24.010,139.083)(24.465,138.621)(24.920,138.146)
  \qbezier(24.920,138.146)(25.375,137.671)(25.830,137.184)
  \qbezier(25.830,137.184)(26.285,136.698)(26.740,136.201)
  \qbezier(26.740,136.201)(27.195,135.705)(27.650,135.200)
  \qbezier(27.650,135.200)(28.105,134.695)(28.560,134.182)
  \qbezier(28.560,134.182)(29.015,133.670)(29.470,133.151)
  \qbezier(29.470,133.151)(29.925,132.632)(30.380,132.107)
  \qbezier(30.380,132.107)(30.835,131.582)(31.290,131.052)
  \qbezier(31.290,131.052)(31.745,130.522)(32.200,129.987)
  \qbezier(32.200,129.987)(32.655,129.452)(33.110,128.913)
  \qbezier(33.110,128.913)(33.565,128.374)(34.020,127.832)
  \qbezier(34.020,127.832)(34.475,127.289)(34.930,126.743)
  \qbezier(34.930,126.743)(35.385,126.197)(35.840,125.648)
  \qbezier(35.840,125.648)(36.295,125.099)(36.750,124.547)
  \qbezier(36.750,124.547)(37.205,123.995)(37.660,123.440)
  \qbezier(37.660,123.440)(38.115,122.886)(38.570,122.329)
  \qbezier(38.570,122.329)(39.025,121.772)(39.480,121.213)
  \qbezier(39.480,121.213)(39.935,120.655)(40.390,120.094)
  \qbezier(40.390,120.094)(40.845,119.533)(41.300,118.970)
  \qbezier(41.300,118.970)(41.755,118.408)(42.210,117.844)
  \qbezier(42.210,117.844)(42.665,117.280)(43.120,116.714)
  \qbezier(43.120,116.714)(43.575,116.149)(44.030,115.582)
  \qbezier(44.030,115.582)(44.485,115.015)(44.940,114.448)
  \qbezier(44.940,114.448)(45.395,113.880)(45.850,113.311)
  \qbezier(45.850,113.311)(46.305,112.742)(46.760,112.172)
  \qbezier(46.760,112.172)(47.215,111.602)(47.670,111.032)
  \qbezier(47.670,111.032)(48.125,110.461)(48.580,109.890)
  \qbezier(48.580,109.890)(49.035,109.318)(49.490,108.746)
  \qbezier(49.490,108.746)(49.945,108.175)(50.400,107.602)
  \qbezier(50.400,107.602)(50.855,107.030)(51.310,106.457)
  \qbezier(51.310,106.457)(51.765,105.884)(52.220,105.311)
  \qbezier(52.220,105.311)(52.675,104.738)(53.130,104.165)
  \qbezier(53.130,104.165)(53.585,103.591)(54.040,103.018)
  \qbezier(54.040,103.018)(54.495,102.444)(54.950,101.871)
  \qbezier(54.950,101.871)(55.405,101.297)(55.860,100.724)
  \qbezier(55.860,100.724)(56.315,100.150)(56.770,99.577)
  \qbezier(56.770,99.577)(57.225,99.003)(57.680,98.430)
  \qbezier(57.680,98.430)(58.135,97.857)(58.590,97.284)
  \qbezier(58.590,97.284)(59.045,96.711)(59.500,96.138)
  \qbezier(59.500,96.138)(59.955,95.566)(60.410,94.993)
  \qbezier(60.410,94.993)(60.865,94.421)(61.320,93.850)
  \qbezier(61.320,93.850)(61.775,93.278)(62.230,92.707)
  \qbezier(62.230,92.707)(62.685,92.136)(63.140,91.565)
  \qbezier(63.140,91.565)(63.595,90.995)(64.050,90.426)
  \qbezier(64.050,90.426)(64.505,89.856)(64.960,89.287)
  \qbezier(64.960,89.287)(65.415,88.719)(65.870,88.151)
  \qbezier(65.870,88.151)(66.325,87.583)(66.780,87.017)
  \qbezier(66.780,87.017)(67.235,86.450)(67.690,85.885)
  \qbezier(67.690,85.885)(68.145,85.319)(68.600,84.755)
  \qbezier(68.600,84.755)(69.055,84.191)(69.510,83.629)
  \qbezier(69.510,83.629)(69.965,83.066)(70.420,82.505)
  \qbezier(70.420,82.505)(70.875,81.944)(71.330,81.384)
  \qbezier(71.330,81.384)(71.785,80.825)(72.240,80.268)
  \qbezier(72.240,80.268)(72.695,79.710)(73.150,79.155)
  \qbezier(73.150,79.155)(73.605,78.599)(74.060,78.046)
  \qbezier(74.060,78.046)(74.515,77.492)(74.970,76.941)
  \qbezier(74.970,76.941)(75.425,76.390)(75.880,75.842)
  \qbezier(75.880,75.842)(76.335,75.293)(76.790,74.748)
  \qbezier(76.790,74.748)(77.245,74.202)(77.700,73.659)
  \qbezier(77.700,73.659)(78.155,73.117)(78.610,72.577)
  \qbezier(78.610,72.577)(79.065,72.038)(79.520,71.502)
  \qbezier(79.520,71.502)(79.975,70.966)(80.430,70.434)
  \qbezier(80.430,70.434)(80.885,69.902)(81.340,69.374)
  \qbezier(81.340,69.374)(81.795,68.846)(82.250,68.323)
  \qbezier(82.250,68.323)(82.705,67.799)(83.160,67.281)
  \qbezier(83.160,67.281)(83.615,66.763)(84.070,66.250)
  \qbezier(84.070,66.250)(84.525,65.737)(84.980,65.230)
  \qbezier(84.980,65.230)(85.435,64.723)(85.890,64.223)
  \qbezier(85.890,64.223)(86.345,63.723)(86.800,63.230)
  \qbezier(86.800,63.230)(87.255,62.738)(87.710,62.253)
  \qbezier(87.710,62.253)(88.165,61.769)(88.620,61.294)
  \qbezier(88.620,61.294)(89.075,60.818)(89.530,60.353)
  \qbezier(89.530,60.353)(89.985,59.889)(90.440,59.436)
  \qbezier(90.440,59.436)(90.895,58.982)(91.350,58.543)
  \qbezier(91.350,58.543)(91.805,58.103)(92.260,57.679)
  \qbezier(92.260,57.679)(92.715,57.254)(93.170,56.848)
  \qbezier(93.170,56.848)(93.625,56.442)(94.080,56.056)
  \qbezier(94.080,56.056)(94.535,55.671)(94.990,55.310)
  \qbezier(94.990,55.310)(95.445,54.949)(95.900,54.618)
  \qbezier(95.900,54.618)(96.355,54.288)(96.810,53.994)
  \qbezier(96.810,53.994)(97.265,53.700)(97.720,53.452)
  \qbezier(97.720,53.452)(98.175,53.204)(98.630,53.016)
  \qbezier(98.630,53.016)(99.085,52.827)(99.540,52.718)
  \qbezier(99.540,52.718)(99.995,52.608)(100.450,52.611)
  \qbezier(100.450,52.611)(100.905,52.611)(101.360,52.781)
  \qbezier(101.360,52.781)(101.815,52.945)(102.270,53.391)
  \qbezier(102.270,53.391)(102.725,53.820)(103.180,54.808)
  \qbezier(103.180,54.808)(103.635,55.696)(104.090,58.199)
  \qbezier(104.090,58.199)(104.545,31.654)(105.000,127.538)
  \put(14.000,127.538){\circle{3.4}}
  \put(105.000,127.538){\circle{3.4}}
  \put(34.266,127.538){\circle*{3.4}}
  \put(60.000,12){\makebox(0,0){\tiny Share of cooperators, $x$}}
  % Panel: alpha=1.0
  \put(173.000,175){\makebox(0,0){\footnotesize $\alpha=1.0$}}
  \put(173.000,161){\makebox(0,0){\tiny Population-proportional}}
  \put(127.000,46.000){\line(0,1){106}}
  \put(127.000,127.538){\line(1,0){91}}
  \put(127.000,46){\line(0,-1){3}}
  \put(127.000,36){\makebox(0,0){\tiny 0}}
  \put(172.500,46){\line(0,-1){3}}
  \put(172.500,36){\makebox(0,0){\tiny 0.5}}
  \put(218.000,46){\line(0,-1){3}}
  \put(218.000,36){\makebox(0,0){\tiny 1}}
  \put(121.000,127.538){\makebox(0,0)[r]{\tiny $0$}}
  \put(127.000,23){\framebox(91,7){}}
  \qbezier(127.000,127.538)(127.455,127.109)(127.910,126.683)
  \qbezier(127.910,126.683)(128.365,126.256)(128.820,125.833)
  \qbezier(128.820,125.833)(129.275,125.410)(129.730,124.990)
  \qbezier(129.730,124.990)(130.185,124.571)(130.640,124.154)
  \qbezier(130.640,124.154)(131.095,123.738)(131.550,123.325)
  \qbezier(131.550,123.325)(132.005,122.912)(132.460,122.503)
  \qbezier(132.460,122.503)(132.915,122.093)(133.370,121.687)
  \qbezier(133.370,121.687)(133.825,121.281)(134.280,120.879)
  \qbezier(134.280,120.879)(134.735,120.477)(135.190,120.079)
  \qbezier(135.190,120.079)(135.645,119.680)(136.100,119.285)
  \qbezier(136.100,119.285)(136.555,118.891)(137.010,118.500)
  \qbezier(137.010,118.500)(137.465,118.109)(137.920,117.722)
  \qbezier(137.920,117.722)(138.375,117.336)(138.830,116.953)
  \qbezier(138.830,116.953)(139.285,116.570)(139.740,116.192)
  \qbezier(139.740,116.192)(140.195,115.813)(140.650,115.438)
  \qbezier(140.650,115.438)(141.105,115.064)(141.560,114.694)
  \qbezier(141.560,114.694)(142.015,114.324)(142.470,113.958)
  \qbezier(142.470,113.958)(142.925,113.592)(143.380,113.231)
  \qbezier(143.380,113.231)(143.835,112.870)(144.290,112.513)
  \qbezier(144.290,112.513)(144.745,112.157)(145.200,111.805)
  \qbezier(145.200,111.805)(145.655,111.453)(146.110,111.106)
  \qbezier(146.110,111.106)(146.565,110.759)(147.020,110.416)
  \qbezier(147.020,110.416)(147.475,110.074)(147.930,109.737)
  \qbezier(147.930,109.737)(148.385,109.400)(148.840,109.067)
  \qbezier(148.840,109.067)(149.295,108.735)(149.750,108.408)
  \qbezier(149.750,108.408)(150.205,108.082)(150.660,107.760)
  \qbezier(150.660,107.760)(151.115,107.438)(151.570,107.122)
  \qbezier(151.570,107.122)(152.025,106.806)(152.480,106.496)
  \qbezier(152.480,106.496)(152.935,106.185)(153.390,105.881)
  \qbezier(153.390,105.881)(153.845,105.576)(154.300,105.277)
  \qbezier(154.300,105.277)(154.755,104.978)(155.210,104.685)
  \qbezier(155.210,104.685)(155.665,104.393)(156.120,104.106)
  \qbezier(156.120,104.106)(156.575,103.819)(157.030,103.539)
  \qbezier(157.030,103.539)(157.485,103.258)(157.940,102.985)
  \qbezier(157.940,102.985)(158.395,102.711)(158.850,102.443)
  \qbezier(158.850,102.443)(159.305,102.176)(159.760,101.915)
  \qbezier(159.760,101.915)(160.215,101.655)(160.670,101.401)
  \qbezier(160.670,101.401)(161.125,101.148)(161.580,100.901)
  \qbezier(161.580,100.901)(162.035,100.655)(162.490,100.415)
  \qbezier(162.490,100.415)(162.945,100.176)(163.400,99.945)
  \qbezier(163.400,99.945)(163.855,99.713)(164.310,99.489)
  \qbezier(164.310,99.489)(164.765,99.265)(165.220,99.049)
  \qbezier(165.220,99.049)(165.675,98.832)(166.130,98.624)
  \qbezier(166.130,98.624)(166.585,98.416)(167.040,98.216)
  \qbezier(167.040,98.216)(167.495,98.017)(167.950,97.825)
  \qbezier(167.950,97.825)(168.405,97.634)(168.860,97.451)
  \qbezier(168.860,97.451)(169.315,97.269)(169.770,97.095)
  \qbezier(169.770,97.095)(170.225,96.922)(170.680,96.757)
  \qbezier(170.680,96.757)(171.135,96.593)(171.590,96.438)
  \qbezier(171.590,96.438)(172.045,96.283)(172.500,96.138)
  \qbezier(172.500,96.138)(172.955,95.993)(173.410,95.858)
  \qbezier(173.410,95.858)(173.865,95.723)(174.320,95.598)
  \qbezier(174.320,95.598)(174.775,95.473)(175.230,95.359)
  \qbezier(175.230,95.359)(175.685,95.245)(176.140,95.141)
  \qbezier(176.140,95.141)(176.595,95.038)(177.050,94.946)
  \qbezier(177.050,94.946)(177.505,94.854)(177.960,94.773)
  \qbezier(177.960,94.773)(178.415,94.693)(178.870,94.624)
  \qbezier(178.870,94.624)(179.325,94.556)(179.780,94.499)
  \qbezier(179.780,94.499)(180.235,94.443)(180.690,94.399)
  \qbezier(180.690,94.399)(181.145,94.355)(181.600,94.324)
  \qbezier(181.600,94.324)(182.055,94.294)(182.510,94.276)
  \qbezier(182.510,94.276)(182.965,94.259)(183.420,94.255)
  \qbezier(183.420,94.255)(183.875,94.252)(184.330,94.263)
  \qbezier(184.330,94.263)(184.785,94.274)(185.240,94.299)
  \qbezier(185.240,94.299)(185.695,94.325)(186.150,94.365)
  \qbezier(186.150,94.365)(186.605,94.406)(187.060,94.462)
  \qbezier(187.060,94.462)(187.515,94.519)(187.970,94.591)
  \qbezier(187.970,94.591)(188.425,94.664)(188.880,94.753)
  \qbezier(188.880,94.753)(189.335,94.843)(189.790,94.950)
  \qbezier(189.790,94.950)(190.245,95.056)(190.700,95.181)
  \qbezier(190.700,95.181)(191.155,95.306)(191.610,95.449)
  \qbezier(191.610,95.449)(192.065,95.592)(192.520,95.755)
  \qbezier(192.520,95.755)(192.975,95.917)(193.430,96.099)
  \qbezier(193.430,96.099)(193.885,96.282)(194.340,96.485)
  \qbezier(194.340,96.485)(194.795,96.687)(195.250,96.912)
  \qbezier(195.250,96.912)(195.705,97.136)(196.160,97.382)
  \qbezier(196.160,97.382)(196.615,97.629)(197.070,97.898)
  \qbezier(197.070,97.898)(197.525,98.167)(197.980,98.460)
  \qbezier(197.980,98.460)(198.435,98.753)(198.890,99.071)
  \qbezier(198.890,99.071)(199.345,99.389)(199.800,99.733)
  \qbezier(199.800,99.733)(200.255,100.076)(200.710,100.446)
  \qbezier(200.710,100.446)(201.165,100.816)(201.620,101.214)
  \qbezier(201.620,101.214)(202.075,101.612)(202.530,102.039)
  \qbezier(202.530,102.039)(202.985,102.466)(203.440,102.922)
  \qbezier(203.440,102.922)(203.895,103.379)(204.350,103.867)
  \qbezier(204.350,103.867)(204.805,104.355)(205.260,104.876)
  \qbezier(205.260,104.876)(205.715,105.396)(206.170,105.951)
  \qbezier(206.170,105.951)(206.625,106.505)(207.080,107.095)
  \qbezier(207.080,107.095)(207.535,107.685)(207.990,108.312)
  \qbezier(207.990,108.312)(208.445,108.939)(208.900,109.604)
  \qbezier(208.900,109.604)(209.355,110.270)(209.810,110.975)
  \qbezier(209.810,110.975)(210.265,111.681)(210.720,112.429)
  \qbezier(210.720,112.429)(211.175,113.177)(211.630,113.969)
  \qbezier(211.630,113.969)(212.085,114.761)(212.540,115.599)
  \qbezier(212.540,115.599)(212.995,116.438)(213.450,117.324)
  \qbezier(213.450,117.324)(213.905,118.211)(214.360,119.148)
  \qbezier(214.360,119.148)(214.815,120.086)(215.270,121.076)
  \qbezier(215.270,121.076)(215.725,122.067)(216.180,123.114)
  \qbezier(216.180,123.114)(216.635,124.160)(217.090,125.266)
  \qbezier(217.090,125.266)(217.545,126.371)(218.000,127.538)
  \put(127.000,127.538){\circle*{3.4}}
  \put(218.000,127.538){\circle{3.4}}
  \put(173.000,12){\makebox(0,0){\tiny Share of cooperators, $x$}}
  % Panel: alpha=2.0
  \put(286.000,175){\makebox(0,0){\footnotesize $\alpha=2.0$}}
  \put(286.000,161){\makebox(0,0){\tiny Popular-biased exposure}}
  \put(240.000,46.000){\line(0,1){106}}
  \put(240.000,127.538){\line(1,0){91}}
  \put(240.000,46){\line(0,-1){3}}
  \put(240.000,36){\makebox(0,0){\tiny 0}}
  \put(285.500,46){\line(0,-1){3}}
  \put(285.500,36){\makebox(0,0){\tiny 0.5}}
  \put(331.000,46){\line(0,-1){3}}
  \put(331.000,36){\makebox(0,0){\tiny 1}}
  \put(234.000,127.538){\makebox(0,0)[r]{\tiny $0$}}
  \put(240.000,23){\framebox(91,7){}}
  \put(306.526,23){\rule{24.474pt}{7pt}}
  \qbezier(240.000,127.538)(240.455,126.859)(240.910,126.185)
  \qbezier(240.910,126.185)(241.365,125.510)(241.820,124.841)
  \qbezier(241.820,124.841)(242.275,124.172)(242.730,123.509)
  \qbezier(242.730,123.509)(243.185,122.846)(243.640,122.189)
  \qbezier(243.640,122.189)(244.095,121.532)(244.550,120.882)
  \qbezier(244.550,120.882)(245.005,120.232)(245.460,119.588)
  \qbezier(245.460,119.588)(245.915,118.944)(246.370,118.308)
  \qbezier(246.370,118.308)(246.825,117.672)(247.280,117.044)
  \qbezier(247.280,117.044)(247.735,116.415)(248.190,115.795)
  \qbezier(248.190,115.795)(248.645,115.175)(249.100,114.563)
  \qbezier(249.100,114.563)(249.555,113.952)(250.010,113.349)
  \qbezier(250.010,113.349)(250.465,112.747)(250.920,112.154)
  \qbezier(250.920,112.154)(251.375,111.561)(251.830,110.979)
  \qbezier(251.830,110.979)(252.285,110.397)(252.740,109.825)
  \qbezier(252.740,109.825)(253.195,109.253)(253.650,108.693)
  \qbezier(253.650,108.693)(254.105,108.133)(254.560,107.585)
  \qbezier(254.560,107.585)(255.015,107.037)(255.470,106.501)
  \qbezier(255.470,106.501)(255.925,105.966)(256.380,105.444)
  \qbezier(256.380,105.444)(256.835,104.922)(257.290,104.414)
  \qbezier(257.290,104.414)(257.745,103.906)(258.200,103.413)
  \qbezier(258.200,103.413)(258.655,102.920)(259.110,102.443)
  \qbezier(259.110,102.443)(259.565,101.966)(260.020,101.505)
  \qbezier(260.020,101.505)(260.475,101.044)(260.930,100.601)
  \qbezier(260.930,100.601)(261.385,100.157)(261.840,99.732)
  \qbezier(261.840,99.732)(262.295,99.307)(262.750,98.901)
  \qbezier(262.750,98.901)(263.205,98.495)(263.660,98.109)
  \qbezier(263.660,98.109)(264.115,97.723)(264.570,97.358)
  \qbezier(264.570,97.358)(265.025,96.993)(265.480,96.649)
  \qbezier(265.480,96.649)(265.935,96.306)(266.390,95.986)
  \qbezier(266.390,95.986)(266.845,95.666)(267.300,95.371)
  \qbezier(267.300,95.371)(267.755,95.075)(268.210,94.804)
  \qbezier(268.210,94.804)(268.665,94.533)(269.120,94.289)
  \qbezier(269.120,94.289)(269.575,94.044)(270.030,93.827)
  \qbezier(270.030,93.827)(270.485,93.610)(270.940,93.421)
  \qbezier(270.940,93.421)(271.395,93.232)(271.850,93.073)
  \qbezier(271.850,93.073)(272.305,92.913)(272.760,92.784)
  \qbezier(272.760,92.784)(273.215,92.656)(273.670,92.559)
  \qbezier(273.670,92.559)(274.125,92.462)(274.580,92.397)
  \qbezier(274.580,92.397)(275.035,92.333)(275.490,92.303)
  \qbezier(275.490,92.303)(275.945,92.272)(276.400,92.277)
  \qbezier(276.400,92.277)(276.855,92.281)(277.310,92.321)
  \qbezier(277.310,92.321)(277.765,92.361)(278.220,92.438)
  \qbezier(278.220,92.438)(278.675,92.514)(279.130,92.629)
  \qbezier(279.130,92.629)(279.585,92.743)(280.040,92.895)
  \qbezier(280.040,92.895)(280.495,93.047)(280.950,93.238)
  \qbezier(280.950,93.238)(281.405,93.429)(281.860,93.660)
  \qbezier(281.860,93.660)(282.315,93.890)(282.770,94.160)
  \qbezier(282.770,94.160)(283.225,94.430)(283.680,94.740)
  \qbezier(283.680,94.740)(284.135,95.050)(284.590,95.399)
  \qbezier(284.590,95.399)(285.045,95.749)(285.500,96.138)
  \qbezier(285.500,96.138)(285.955,96.528)(286.410,96.956)
  \qbezier(286.410,96.956)(286.865,97.385)(287.320,97.852)
  \qbezier(287.320,97.852)(287.775,98.319)(288.230,98.825)
  \qbezier(288.230,98.825)(288.685,99.330)(289.140,99.872)
  \qbezier(289.140,99.872)(289.595,100.414)(290.050,100.991)
  \qbezier(290.050,100.991)(290.505,101.568)(290.960,102.179)
  \qbezier(290.960,102.179)(291.415,102.791)(291.870,103.433)
  \qbezier(291.870,103.433)(292.325,104.076)(292.780,104.749)
  \qbezier(292.780,104.749)(293.235,105.421)(293.690,106.121)
  \qbezier(293.690,106.121)(294.145,106.820)(294.600,107.544)
  \qbezier(294.600,107.544)(295.055,108.268)(295.510,109.013)
  \qbezier(295.510,109.013)(295.965,109.758)(296.420,110.521)
  \qbezier(296.420,110.521)(296.875,111.283)(297.330,112.060)
  \qbezier(297.330,112.060)(297.785,112.837)(298.240,113.624)
  \qbezier(298.240,113.624)(298.695,114.411)(299.150,115.204)
  \qbezier(299.150,115.204)(299.605,115.998)(300.060,116.793)
  \qbezier(300.060,116.793)(300.515,117.588)(300.970,118.381)
  \qbezier(300.970,118.381)(301.425,119.174)(301.880,119.960)
  \qbezier(301.880,119.960)(302.335,120.745)(302.790,121.520)
  \qbezier(302.790,121.520)(303.245,122.294)(303.700,123.052)
  \qbezier(303.700,123.052)(304.155,123.811)(304.610,124.548)
  \qbezier(304.610,124.548)(305.065,125.286)(305.520,125.998)
  \qbezier(305.520,125.998)(305.975,126.711)(306.430,127.394)
  \qbezier(306.430,127.394)(306.885,128.077)(307.340,128.727)
  \qbezier(307.340,128.727)(307.795,129.377)(308.250,129.989)
  \qbezier(308.250,129.989)(308.705,130.601)(309.160,131.172)
  \qbezier(309.160,131.172)(309.615,131.743)(310.070,132.269)
  \qbezier(310.070,132.269)(310.525,132.795)(310.980,133.274)
  \qbezier(310.980,133.274)(311.435,133.752)(311.890,134.180)
  \qbezier(311.890,134.180)(312.345,134.608)(312.800,134.983)
  \qbezier(312.800,134.983)(313.255,135.358)(313.710,135.678)
  \qbezier(313.710,135.678)(314.165,135.998)(314.620,136.261)
  \qbezier(314.620,136.261)(315.075,136.525)(315.530,136.730)
  \qbezier(315.530,136.730)(315.985,136.936)(316.440,137.083)
  \qbezier(316.440,137.083)(316.895,137.229)(317.350,137.317)
  \qbezier(317.350,137.317)(317.805,137.404)(318.260,137.432)
  \qbezier(318.260,137.432)(318.715,137.460)(319.170,137.429)
  \qbezier(319.170,137.429)(319.625,137.398)(320.080,137.308)
  \qbezier(320.080,137.308)(320.535,137.217)(320.990,137.069)
  \qbezier(320.990,137.069)(321.445,136.921)(321.900,136.716)
  \qbezier(321.900,136.716)(322.355,136.510)(322.810,136.249)
  \qbezier(322.810,136.249)(323.265,135.988)(323.720,135.673)
  \qbezier(323.720,135.673)(324.175,135.358)(324.630,134.990)
  \qbezier(324.630,134.990)(325.085,134.622)(325.540,134.204)
  \qbezier(325.540,134.204)(325.995,133.785)(326.450,133.318)
  \qbezier(326.450,133.318)(326.905,132.850)(327.360,132.336)
  \qbezier(327.360,132.336)(327.815,131.822)(328.270,131.263)
  \qbezier(328.270,131.263)(328.725,130.704)(329.180,130.103)
  \qbezier(329.180,130.103)(329.635,129.501)(330.090,128.860)
  \qbezier(330.090,128.860)(330.545,128.218)(331.000,127.538)
  \put(240.000,127.538){\circle*{3.4}}
  \put(331.000,127.538){\circle*{3.4}}
  \put(306.526,127.538){\circle{3.4}}
  \put(286.000,12){\makebox(0,0){\tiny Share of cooperators, $x$}}
\end{picture}
\caption{The curves show $\dot x=x^\alpha/[x^\alpha+e(1-x)^\alpha]-x$. The black segments in the bars beneath each panel mark regions where effective MLRP fails; unshaded segments mark regions where it holds. Filled dots denote stable rest points; open dots denote unstable rest points.}
\label{fig:pd-mlrp}
\end{figure}

To illustrate Corollary~1, consider the prisoner's-dilemma game in
Table~\ref{tab:pd-example}. Agents are uniformly randomly matched in pairs
and play the game. Let $x\in[0,1]$ denote the population share of cooperators.
The expected payoffs generated by this random matching are
\[
F_C(x)=3x+0(1-x)=3x,
\qquad
F_D(x)=4x+1(1-x)=1+3x.
\]
Thus, defection yields a payoff advantage of one at every population state.
We use the good-news probability $a(u)=ce^u$, where $c>0$ is chosen sufficiently
small that $a(u)<1$ for all feasible payoffs, and the power exposure function
$m_i(x)=x_i^\alpha/\sum_j x_j^\alpha$. Figure~\ref{fig:pd-mlrp} illustrates how exposure biases alter
evolutionary stability and identifies the regions where effective
MLRP holds or fails. Under population-proportional exposure,
the unique Nash equilibrium $x=0$ is asymptotically stable.
Rare-biased exposure destabilizes $x=0$, whereas popular-biased
exposure makes the non-Nash pure state $x=1$ asymptotically
stable while preserving the stability of $x=0$.

\begin{table}[htbp]
\centering
\begin{tabular}{c|c|c|}
\multicolumn{1}{c}{} &
\multicolumn{1}{c}{$C$} &
\multicolumn{1}{c}{$D$} \\
\cline{2-3}
$C$ & $3$ & $0$ \\
\cline{2-3}
$D$ & $4$ & $1$ \\
\cline{2-3}
\end{tabular}
\caption{Prisoner's dilemma.}
\label{tab:pd-example}
\end{table}

\enlargethispage{2\baselineskip}

\clearpage
\section{Conclusion}

This note examines how the interaction between payoff information and strategy exposure shapes evolutionary dynamics. We distinguish MLRP of the original payoff signals from effective MLRP, which determines whether informational advantages translate into faster relative population growth. Our results show that even when better-performing strategies are more likely to generate good news, exposure biases can overturn their evolutionary advantage. Sufficiently strong exposure favoring rare strategies can destabilize strict Nash equilibria, while exposure favoring popular strategies can stabilize non-Nash pure states. Thus, holding the underlying game and information structure fixed, different exposure mechanisms can generate sharply different evolutionary predictions.

\subsection*{Declaration of AI use}
OpenAI's ChatGPT was used during the development of this study to discuss and refine ideas, improve exposition, assist with editing and \LaTeX\ formatting, prepare figures, and check the presentation of mathematical arguments. The research questions, the model and the results are the authors' own. The authors checked every proof and all AI-assisted material and take full responsibility for the final manuscript.
\bibliographystyle{ecta}
\bibliography{bib}

\end{document}